\documentclass[a4paper,12pt]{article}
\usepackage[russian]{babel}
\usepackage{amsmath}
\usepackage{ulem}
\usepackage{calc}
\usepackage{vmargin}
\usepackage{amstext}
\usepackage{graphicx}
\usepackage{amsfonts}
\usepackage{amssymb}
\usepackage{titlesec}

\numberwithin{equation}{section}

\begin{document}

\title{\bf Solutions of the Euler equations and stationary structures in an inviscid fluid. }

\author{\bf O. V.  Kaptsov\\ 
	Institute of Computational Modeling, Siberian Branch,\\
	Russian Academy of Sciences, Krasnoyarsk, Russia\\
	E-mail: kaptsov@icm.krasn.ru
	 }
\date{}
\maketitle

{\bf Abstract.}

The Euler equations describing two-dimensional steady flows of an  inviscid fluid are studied.
These equations are reduced to one equation for the stream function and then, using the Hirota function, solutions of three nonlinear elliptic equations are found.
The solutions found are interpreted as sources in a rotating fluid,
jets, chains of sources and sinks, vortex structures.
We propose a new simple method for constructing solutions in the form of rational expressions of elliptic functions.
It is shown that the flux of fluid across a closed curve is quantized
in the case of the elliptic  Sin-Gordon equation. 

\noindent
{\it Keywords:} Euler equations, Hirota function, elliptic functions. 

\section{Introduction}

It is well known that the equations   
$$ uu_x+vu_y+p_x=0 ,\qquad   uv_x+vv_y+p_y=0 ,\qquad u_x+v_y=0 ,$$
describe  steady two-dimensional flows of inviscid fluid,   
where $u, v$ are the components of velocity and $p$ is the pressure.
This system is reduced to one equation  
 \begin{equation} \label{psi} 
\Delta \psi = \omega(\psi) 
\end{equation}	
for the stream function $\psi$ \cite{Batchelor}. 
Here, and below,  $\Delta$ represents the two-dimensional Laplacian operator, 
$\omega$ is vorticity and $u=\psi_y$, $v=-\psi_x$.	
Equation (\ref{psi}) also occurs in various applications such as plasma physics and condensed
matter physics \cite{Bateman,Movsesyants,Borisov}.

Equation (\ref{psi}) is well investigated in the linear case.
In addition, the elliptic Liouville equation
$$\Delta \psi = \exp(\psi) $$
connected by transformation
$$\psi= \log(-2\Delta(\log\tau)) $$
with the Laplace equation $\Delta \tau=0$.	
This is a direct consequence of the classical formula for the general solution of the hyperbolic Liouville equation \cite{Ibragimov}.

At the end of the nineteenth and beginning of the twentieth century, the B\"{a}cklund and  Tzitzeica \cite{Rogers}
found transformations that allow generating solutions to the equations
$$ u_{xy} = \sin(u) , \qquad u_{xy} = \exp(u) - \exp(-2u) . $$
 Using the  B\"{a}cklund transformation, in \cite{Borisov} 
some vortex-type singular solutions of the elliptic Sine-Gordon equation
 \begin{equation} \label{SG} 
 	\Delta \psi = \sin(\psi)  
 \end{equation} 
    were found.
Multiparameter solution formula for the Tzitzeica equation
$$ \Delta \psi = \exp(\psi) - \exp(-2\psi) $$
was presented in \cite{KaptsovTzi}.
Note the  works \cite{Sher,KaptsovBook}, which used separation of variables to construct solutions of the equation (\ref{psi}) with other functions $\omega(\psi)$.

This paper is divided into two parts. The first deals with solutions of the Sine-Gordon equation (\ref{SG}) and Sinh-Gordon one.
So the solutions of the Sine-Gordon equation are represented as
$$	\psi =	4\tan^{-1} \frac{G}{F} \ ,   $$
where $F$ and $G$ are smooth functions in the plane $\mathbb{R}^2(x,y)$.
It is shown that the flux of fluid volume 
crossing the simple closed curve $\gamma\subset \mathbb{R}^2(x,y)$ is equal to 
$$ Q\equiv \oint_{\gamma} d\psi = 8\pi I ,\qquad  I\in \mathbb{Z}.$$
 The integer $I$ is equal to the sum of the Poincaré indices of zero points of the vector field $V=(F,G)$ lying inside the bounded curve $\gamma$.
 We have found exact solutions of the Sine-Gordon (\ref{SG}) and Sinh-Gordon equations expressed in terms of elementary functions.
 These solutions can be interpreted as sources and sinks, jet streams, periodic chains of sources and sinks, vortexes, and combinations thereof.
 
In the second part, a new method for constructing elliptic solutions of the Sin-Gordon, Sinh-Gordon and Tzitz\'{e}ica equations is proposed. These classes of solutions are represented as rational expressions of elliptic functions.	To find them, the Maple computer algebra system is used.
The calculations are similar to those performed in \cite{K+K}.

\section{Elementary solutions}

In this section we will find some elementary solutions of equation  (\ref{psi}), that is 
solutions which can be expressed in terms of algebraic operations, logarithms and exponentials.
We begin with the Sine-Gordon and  Sinh-Gordon equations
\begin{equation} \label{sin} 
\Delta\psi =\sin(\psi), 
\end{equation}	
\begin{equation} \label{sinh} 
\Delta\psi =\sinh(\psi) . 
\end{equation}
Let us reduce  these equations to a same form using complex and double numbers \cite{Olariu}.
The field of complex numbers will be denoted by $\mathbb{C}$, and the algebra of double numbers by $\mathbb{D}$. Every complex and double number has the form $z=a+\delta b$, where $a, b\in \mathbb{R}$, $\delta\notin\mathbb{R}$. 
So if $\delta=i\in \mathbb{C}$, then $\delta^2$  is equal to $-1$, and if $\delta\in \mathbb{D}$, then $\delta^2 $ is equal to $1$. Multiplication of double numbers is given by:
$$ (a_1+\delta b_1) (a_2+\delta b_2) =a_1a_2+b_1b_2+\delta (a_1 b_2+a_2b_1) .      $$
Thus, the equations (\ref{sin}), (\ref{sinh}) can be written as
$$\Delta\psi = \frac{\exp(\delta \psi)- \exp(-\delta \psi)}{2 \delta} .   $$
Let $v=\exp(\delta \psi)$ be a new function,  then the previous equation is of the form
\begin{equation} \label{eq_v} 
	v(v_{xx} +v_{yy}) -v^2_x-v^2_y -v^3/2+v/2 = 0 .
\end{equation}

Suppose $F$ and $G$ are smooth functions on an open set $\Omega\subset \mathbb{R}^2$ and  $H = F+\delta G$.  Then we say that the function $\bar{H} =F-\delta G$ is conjugate to  $H$.
Next we  look for solutions of (\ref{sin}), (\ref{sinh}) in the form
 \begin{equation} \label{v} 
	v =\frac{\bar{H}^2}{H^2} .
\end{equation}
Obviously, the functions $v$ and $\psi$ can be written as
$$  v=\left(\frac{1-\delta\frac{G}{F}}{1+\delta\frac{G}{F}} \right)^2 ,\qquad
  \psi =  \frac{2}{\delta}  \log\left(\frac{1-\delta\frac{G}{F}}{1+\delta\frac{G}{F}} \right)  .        $$
Therefore the function  $\psi$ is 
\begin{equation} \label{arctan} 
\psi =	4\tan^{-1} \frac{G}{F} \ ,  
\end{equation}
when $\delta = i\in\mathbb{C}$; but if  $\delta \in \mathbb{D}$, then 
$$\psi =	4 \tanh^{-1} \frac{G}{F}  \ . $$

We note the useful statement about  the amount of fluid  crossing a closed curve.
Suppose the stream function $\psi$ satisfies the  Sine-Gordon equation and has the form (\ref{arctan}), where $F$ and $G$ generate a vector field $V=(F,G)$ in the domain  $\Omega\subset\mathbb{R}^2$. Let $a\in\Omega$ be an zero of $V$ and 
$c$ is a small circle around the zero.
Then the line integral
$$ind(a) \equiv \frac{1}{2\pi}\oint_{c} d\tan^{-1}(G/F)    $$
is integer  called Poincar\'{e}'s index of the point $a$ \cite{Andronov}.
Thus,  the source strength
$$ \oint_{c} d\psi , $$  
is equal to $8\pi ind(a)\in 8\pi\mathbb{Z}$, i.e.,  the source strength is quantized  in this case.
We say that a line integral
$$ q=\frac{1}{2\pi}\oint_{c} d\psi  $$ 
is the topological charge of the point $a$.

Let us suppose that  a simple closed curve $\gamma$ is the boundary of $\Omega$ and 
 the vector field $V$ has several zeros $a_1,\dots a_n$ with  topological charges $q_1,\dots,q_n$.
 It follows from  Poincar\'{e}'s theorem \cite{Arnold} that
$$\oint_{\gamma} d\psi  =  8\pi\sum_{j=1}^{n} ind(a_j)= 2\pi\sum_{j=1}^{n} q_j . $$
 Thus,  the flux of fluid volume across the closed curve $\gamma$ is  
 $$ Q\equiv\oint_{\gamma} d\psi = 2\pi\sum_{j=1}^{n} q_j  \in 8\pi\mathbb{Z},$$
 i.e., this means that $Q$ is also quantized.   
 A useful way to think of  singular solutions with topological charges is as  point defects in the fluid.
Other topological quantum numbers are discussed in \cite{Thouless}.



  We now look for solutions of the equations (\ref{sin}) and (\ref{sinh}) by using the function $H$ of the form
$$  1 + \delta\exp(kx+ny+\eta) , $$
with  $k, n,\eta\in\mathbb{R}$.
It is easy to see that $v$ satisfies the equation (\ref{eq_v}), if $k^2+n^2=1$.
In the case of the Sine-Gordon equation, the function $\psi$ is smooth; its graph is a two-dimensional kink. 
The streamlines $\psi=const$ corresponding to this solution are obviously straight lines.
 For simplicity we assume that $k=1, n=\eta=0$. The corresponding stream function and  velocity components are
 $$ \psi =	4 \tanh^{-1} (\exp(x)) ,\qquad v_x=0, \quad 
  v_y= -\frac{2}{\cosh^2(x/2)} . $$
 Then the solution may be interpreted as a steady jet that is parallel to the y-axis. When 
 $kn\neq 0$ we also have a jet flow. 
We remark that in the case of the Sinh-Gordon equation, the solution $\psi$  is discontinuous.

Next, we call any linear combination of the functions 
$\exp(kx+ny+\eta)$, with $k, n,\eta\in\mathbb{C}$, a Hirota function.
Consider a Hirota function
\begin{equation} \label{tau2} 
	H = 1 + \delta (f_1+f_2) + \delta^2 s_{12} f_1f_2 ,
\end{equation}
where $f_i =\exp(k_ix+n_iy+\eta_i)$, $i=1,2 $. Substituting the function (\ref{v}) into left side of  (\ref{eq_v}), we obtain a rational function whose numerator is a polynomial $P$ in  $k_i,n_i,s_{12}$. Equating the coefficients of $P$ to zero, we have a system NAS of nonlinear algebraic equations. The system has a non-trivial solution
$$ s_{12} = \frac{n_1n_2+k_1k_2-1}{n_1n_2+k_1k_2+1} , \qquad n_i^2+k_i^2=1 \quad i=1,2  .   $$
Hence solutions of the equations (\ref{sin}), (\ref{sinh}) are given by
$$   \psi_1 =4\tan^{-1} \left(\frac{f_1+f_2}{1-s_{12}f_1f_2}\right) , \qquad  
\psi_2 =4\tanh^{-1}\left( \frac{f_1+f_2}{1+s_{12}f_1f_2}\right)   .                       $$

\begin{figure}  
	\centering
	\includegraphics[height=10cm, width=10cm]{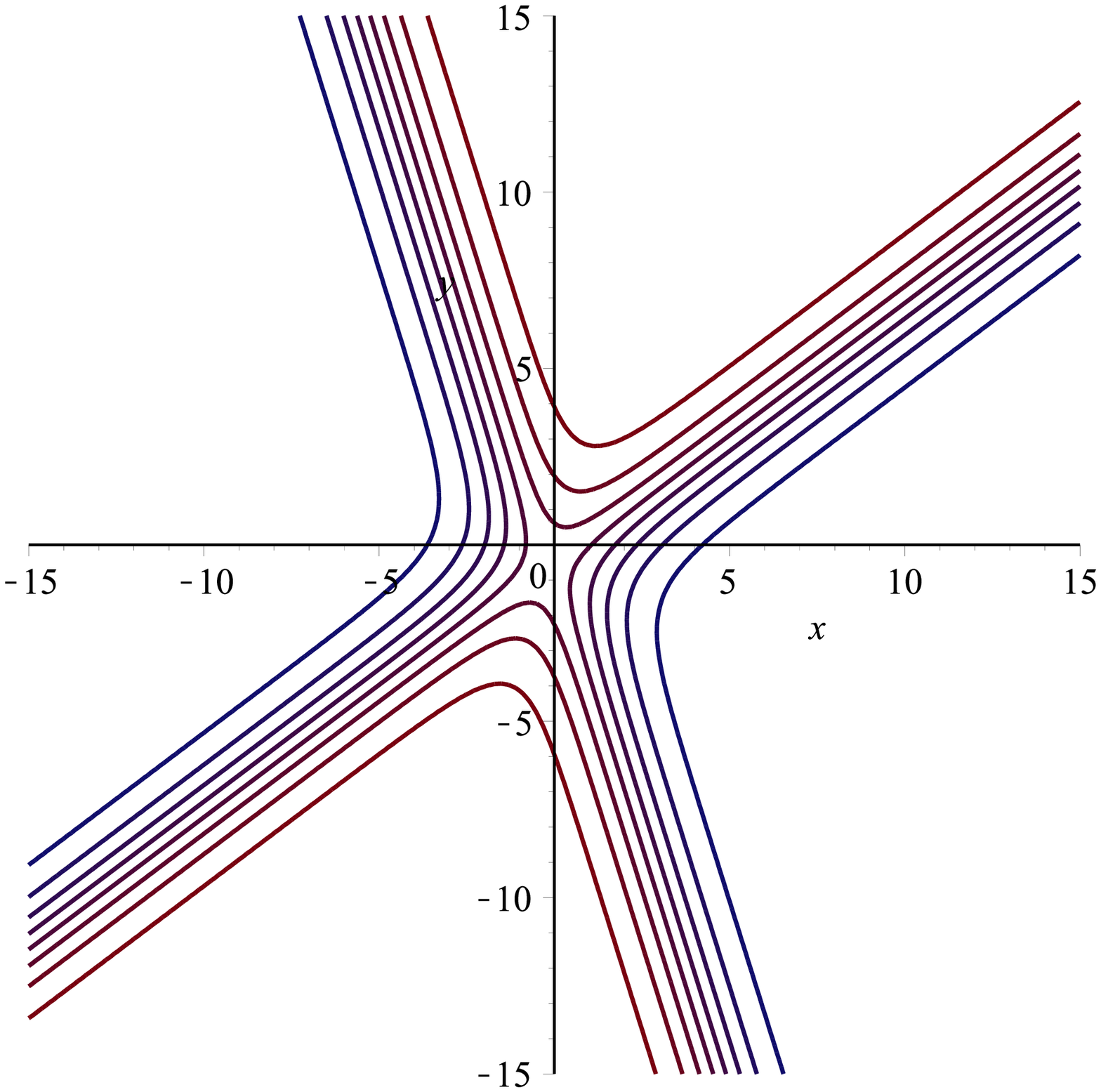}
	\caption{two jets.} 
		\label{fig:1}
\end{figure}

We further consider only the solution $\psi_1$, since the function  $\psi_2$ is discontinuous. 
Suppose that $\eta_1=\eta_2=0$, $n_i, k_i\in\mathbb{R}$. In this case, the graph of $\psi_1$ 
looks like two kinks and has one saddle point. Streamlines are shown in Figure \ref{fig:1}. 
It may be interpreted as an interaction of two jets. 

\begin{figure}	
 	\centering
 	\includegraphics[height=10cm, width=10cm]{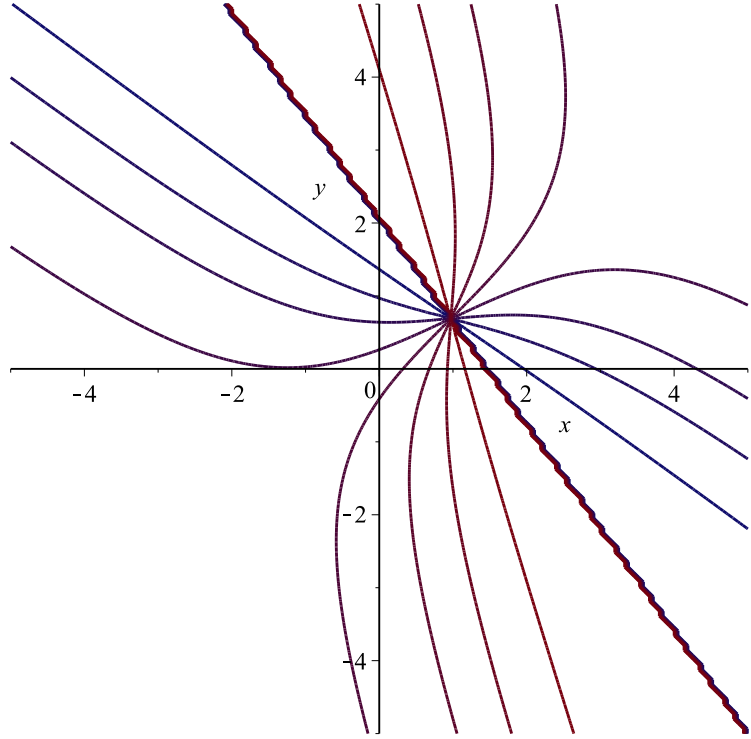}
 	\caption{source or sink.}	\label{fig:img2}
 \end{figure}

Now we set $\eta_1=0$, $\eta_2=i\pi$, $n_i, k_i\in\mathbb{R}$. 
It gives a singularity of the velocity distribution at some point $A$, where $G=f_1+f_2=0$ and  $F=1-s_{12}f_1f_2=0$. 
As we said above, this solution may be interpreted as a point source or sink in the rotational flow.
 Because of $A$ is a simple point of vector field $V=(F,G)$ then the topological charge of the point $A$ is equal to $\pm 4$ and the source strength is $\pm 8\pi$. Since equations (\ref{sin}) and (\ref{sinh}) are invariant under the transformation  $\psi \longrightarrow -\psi$ we can obtain a source or a sink. Figure \ref{fig:img2} shows the pattern of streamlines in  the $(x,y)$-plane for the flow associated with a source or a sink. 

Let us set $\eta_1 = \eta_2 = 0$, $n_1=a+ib, n_2=a-ib$ ($b\neq 0$). Then the relation  $f_1+f_2=0$ gives countable number of straight and parallel  streamlines. The perpendicular line defined by the equation 
$ 1 - s_{12} f_1 f_2 = 0$ 
is also a streamline. 
The points of intersection of the streamlines are  sources or sinks.
Each semi-infinite strip between the nearest parallel streamlines is filled with streamlines connecting a source and a sink.
  
  Let us consider a Hirota function
\begin{equation} \label{tau3} 
	\tau = 1 + \delta (f_1+f_2+f_3) + \delta^2 (s_{12} f_1f_2 +s_{13} f_1f_3+s_{23} f_2f_3 )+ \delta^3 s_{123}f_1f_2f_3 .
\end{equation}
Here, as above, the functions $f_i$ have the form $\exp(k_ix+n_iy+\eta_i)$,  with $k_i,n_i,\eta_i\in\mathbb{C}$, and $s_{ij}$ is given by the following formula
\begin{equation} \label{Sij}
 s_{ij} = \frac{n_in_j+k_ik_j-1}{n_in_j+k_ik_j+1} , \qquad n_i^2+k_i^2=1   \quad (i=1,2,3).  
\end{equation}
Substituting the Hirota function (\ref{tau3}) into the equation (\ref{eq_v}) and (\ref{v}), we find 
$$s_{123}=s_{12}s_{13}s_{23}.     $$

\begin{figure}[h!]	
	\centering
	\includegraphics[height=10cm, width=10cm]{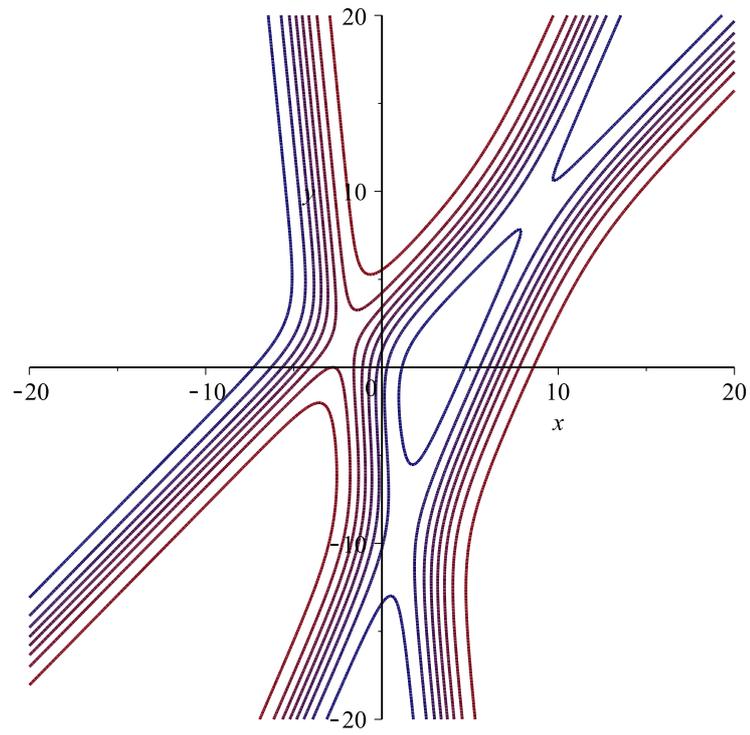}
		\caption{three jets and a vortex.}
			\label{fig:3}
\end{figure}

To obtain flow patterns, one must again choose constants $n_i, \eta_i\in\mathbb{C}$ $(i=1,2,3)$ and  signs of $k_i=\pm \sqrt{1-n_i^2} $. We have the simplest case when $\eta_i=0$ and $n_i, k_i\in \mathbb{R}\ (i=1,2,3)$.
If $n_1=-0.7, n_2=0.4, n_3=0.1$, $k_1=\sqrt{1-k_1^2}, k_2 = - \sqrt{1-k_2^2}, k_3=\sqrt{1- k_3^2}$, then we get the flow pattern shown in Figure \ref{fig:3}. There we see three jets and a vortex.

\begin{figure}[h!]	
	\centering
	\includegraphics[height=10cm, width=10cm]{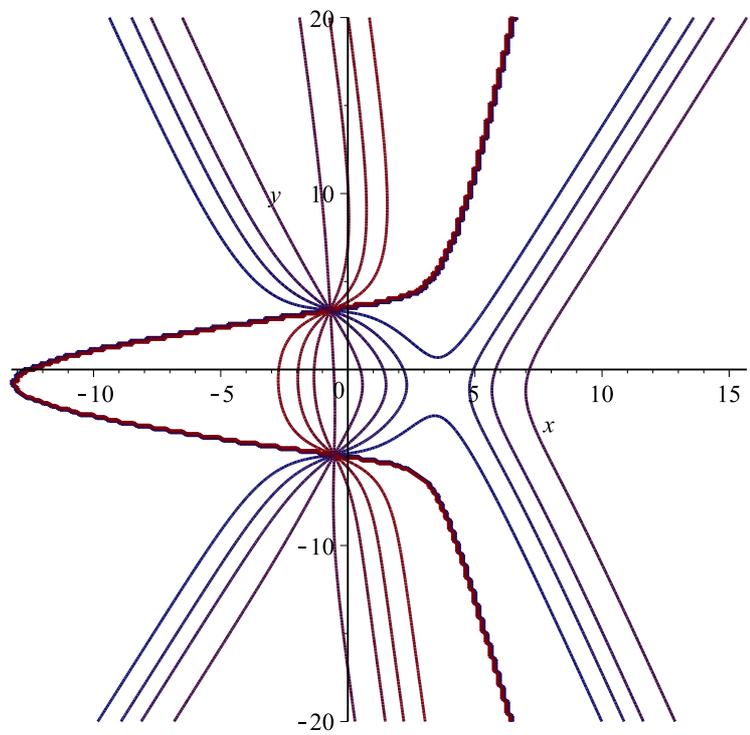}
	\caption{ a sink, a source and jets.}
	\label{fig:4}
\end{figure}

If we set $\eta_1=i\pi, \eta_2=\eta_3=0$, $n_1=0.1, n_2=0.4, n_3=0.3$ and $k_1=\sqrt{1-k_1^2}$, $k_2 = - \sqrt{1-k_2^2}, k_3=\sqrt{1-k_3^2}$, then we have the flow pattern, including a sink, a source and jets (see Fig. \ref{fig:4}). It is easy to construct other solutions by choosing, for example, $n_1$ and $n_2$ to be complex conjugate and $n_3$ to be real.

Formulas (\ref{tau2}) and (\ref{tau3}) can be written as
$$   (1+\delta f_1)*(1+\delta f_2)  ,\qquad  (1+\delta f_1)*(1+\delta f_2)*(1+\delta f_3) . $$
Here addition is defined in the usual way, and the operation $*$ is given by the formulas
$$ f_i*f_j=s_{ij}f_if_j, \qquad f_i* f_j*f_k= s_{ij}s_{ik}s_{jk}f_if_jf_k , $$
where $s_{ij}, s_{ik}, s_{jk}$ are calculated according to (\ref{Sij}).
For arbitrary  $n$ the Hirota function  has the form of the product
$$ H_n = (1+\delta f_1)*(1+\delta f_2)*\dots *(1+\delta f_{n-1})*(1+\delta f_n) .  $$
In this case, the following relations must be fulfilled
$$ f_{i_1}*\dots *f_{i_m} = s_{i_1\dots i_m} f_{i_1}\cdots f_{i_m} , $$
where $s_{i_1\dots i_m}$ is the product of all $s_{jk}$ such that $j,k\in \{i_1,\dots i_m\}$ and
$j<k$.

\begin{figure}[h!]
	\centering
	\includegraphics[height=10cm, width=10cm]{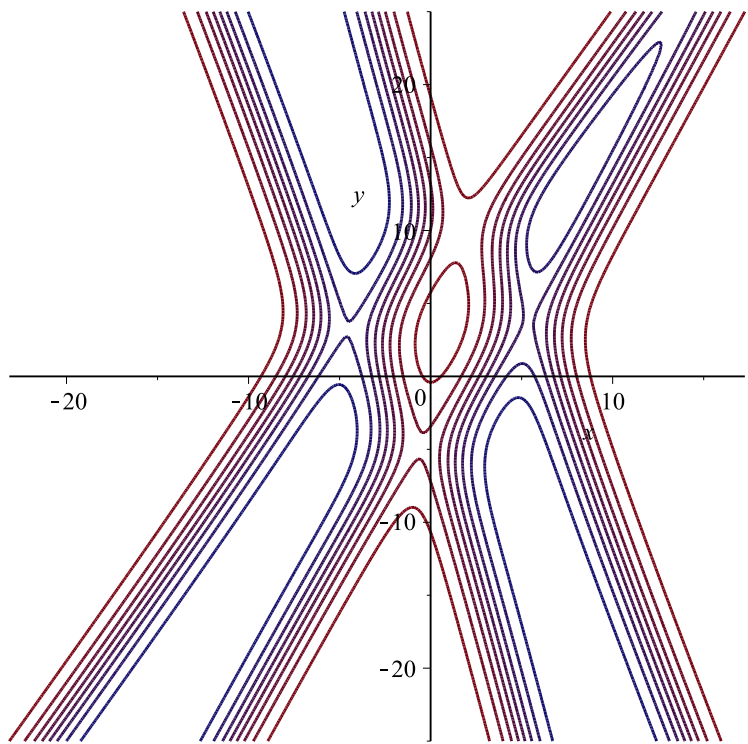}
	\caption{four jets and two vortices.}
		\label{fig:5}
\end{figure}

Let us consider only some solutions to the Sine-Gordon equation for the case $n=4$. The corresponding Hirota function is 
$$ H_4 =  (1+\delta f_1)*(1+\delta f_2)*(1+\delta f_3)*(1+\delta f_4) .  $$
We set $\eta_i=0$ $(1\leq i\leq 4)$ and $ n_1 = 0.5, n_2 = 0.4, n_3 = 0.3, n_4 = 0.2$, when
$k_i<0$ $(1\leq i\leq 3)$, $k_4>0$. This gives a flow pattern with four jets and two vortices (see Fig. \ref{fig:5}).
Assuming one of $\eta_i$ is imaginary, we can get a pair of sources and a pair of sinks or a singe sink.
Choosing $n_1=a+ib$, $n_2=a-ib$, $n_3=a+ib$ and $n_4=a-ib$ we obtain intersecting
chains of sources and sinks, or the resonance of these chains resembling the resonance of solitons for the  Kadomtsev-Petviashvili equation.

At the end of this section a few words should be said about the solutions of the Tzitz\'{e}ica equation
\begin{equation} \label{tsi} 
\Delta	\psi = \exp(\psi)-\exp(-2\psi) .
\end{equation}

We have proposed in \cite{KaptsovTzi} the following representation  
$$ \psi= \log\left(1-2(\Delta\log H)  \right)     $$
  of solutions of the Tzitzeica equation.   The function $\psi$ satisfies the equation (\ref{tsi}) if $H$ is the Hirota function
  $$ H = (1+ f_1)*(1+ f_2)*\dots *(1+ f_n) ,  $$
  where   $f_i=\exp(k_ix+n_iy+\eta_i) $,\ $n_i^2+k_i^2=3$,\  $f_i*f_j=s_{ij}f_if_j$, 
$$  s_{ij}=\frac{4k_ik_jn_in_j+4k_i^2k_j^2-9n_in_j-6k_i^2-9k_ik_j-6k_j^2  +27}{4k_ik_jn_in_j+4k_i^2k_j^2+9n_in_j-6k_i^2+9k_ik_j-6k_j^2  +27} .$$

\section{Solutions expressed in elliptic functions}

The Sine-Gordon equation (\ref{sin}) has exact solutions that are expressed in terms of elliptic functions. Some such solutions were found using the Bäcklund transformation \cite{Borisov} and the Steuerwald ansatz \cite{Ste,Kap19B8}, often called the Lamb ansatz.
Unfortunately, the monograph \cite{KaptsovBook} contains some errors in constructing solutions in elliptic functions. In this section we will find a number of new ansatz for the equations (\ref{sin}) and (\ref{tsi}).
It should be noted that in the case of hyperbolic Sine-Gordon equation, solutions expressed in terms of elliptic functions were obtained in \cite{Matveev} using reductions from algebro-geometric solutions.

 Let us now introduce a new function 
$ w=     \tan(\psi/4). $
Then the equation (\ref{sin}) is rewritten as
\begin{equation} \label{eqW}
	(1+w^2)(w_{xx}+w_{yy}) - 2w(w_x^2+w_y^2)+w^3-w = 0.
\end{equation}
We look for solutions of the last equation in the form
\begin{equation} \label{w}
	w= \frac{s_0+s_1F+s_2G}{p_0+p_1F+p_2G} ,
\end{equation}
with $s_i, p_i\in\mathbb{R}$ $(i=0,1,2)$. Assume that $F(x), G(y)$ are functions satisfying ordinary differential equations
\begin{equation} \label{Fprime}
	(F_x^{\prime})^2= a_4F^4 +a_3F^3 +a_2F^2+a_1F +a_0 ,\qquad a_i\in\mathbb{R},
\end{equation}
\begin{equation} \label{Gprime}
	(G_y^{\prime})^2= b_4G^4 +b_3G^3 +b_2G^2+b_1G +b_0 ,\qquad b_i\in\mathbb{R}.
\end{equation}
Substitute (\ref{w}) into the left side of the equation (\ref{eqW}) and express all derivatives using (\ref{Fprime}) and (\ref{Gprime}). As a result, we obtain a rational function of $F$ and $G$.
Equating the coefficients of the numerator to zero, we obtain a NAS system of nonlinear algebraic equations with respect to $a_i, b_i$ ($0\leq i\leq 4$) and $s_j, p_j$ ($0\leq j\leq 2$ ).
NAS system solutions are quite cumbersome and therefore we consider only two cases.

The first case is an analogue of the Steuerwald ansatz
 \begin{equation} \label{Ste} 
	w = \frac{G}{F} , 
\end{equation} 
where the functions $F, G$ satisfy the equations
$$	(F_x^{\prime})^2= b_4F^4 +(1-b_2)F^2+b_0 ,\quad (G_y^{\prime})^2= b_4G^4 +b_2G^2+b_0 ,
\quad b_0, b_2, b_4\in\mathbb{R} .
$$
If $b_4<0$, then the functions are expressed in terms of the Jacobi function $dn$ (the delta amplitude). Thus the functions $F$ and $G$ are periodic and vanish twice on the period.
As a result, we have a partition of $xy$-plane into equal rectangles.
There are sources (or sinks, respectively) at two opposite vertices of the rectangle; inside there is one saddle point, and streamlines connect sources and sinks.

The second solution has the form
 \begin{equation} \label{w2} 
 	w = \frac{F-G}{F+G} .
 \end{equation} 

Moreover, the functions $F(x)$ , $G(y)$ satisfy the equations
$$	(F_x^{\prime})^2= b_4F^4 -(1+b_2)F^2+b_0 ,\qquad (G_y^{\prime})^2= b_4G^4 +b_2G^2+b_0 .  $$
The flow pattern is qualitatively the same as in the previous case.

Now we look for solutions to equation (\ref{eqW}) in the form
\begin{equation} \label{w3} 
	w= \frac{s_0+s_1F+s_2G+s_3FG}{p_0+p_1F+p_2G+p_3FG} \ . 
\end{equation}
Assume that the functions $F, G$ satisfy equations (\ref{Fprime}), (\ref{Gprime}).
Substitute the $w$ given by the formula (\ref{w3}) into the left side of (\ref{eqW}) and express all derivatives using (\ref{Fprime}), (\ref{Gprime}).
Then we have a rational function of $F, G$, and equating the numerator coefficients to zero, we obtain a nonlinear algebraic system with respect to $a_i, b_i$ ($0\leq i\leq 4$) and $s_j, p_j$ ($0\leq j\leq 3$ ). Its solutions can be found using computer algebra systems. Here are some representations for the function $w$:
$$ w = \frac{s_0+s_1F+s_0s_3G/s_1+s_3FG}{p_2G}  \ ,\qquad w = \frac{s_1F +s_3FG}{p_2G+p_3FG}\ , $$
$$ w = 4\frac{a_0s_1F +s_3FG}{p_0(4a_0 + a_1 F)} \ ,  \qquad
 w = \frac{s_1s_2/s_3 +s_1F+s_2G+s_3FG}{p_0} \ . $$
 Coefficients of ordinary differential equations for the functions $F, G$ are cumbersome and we do not present them.
 
\begin{figure}[h!]
 	\centering
 	\includegraphics[height=10cm, width=10cm]{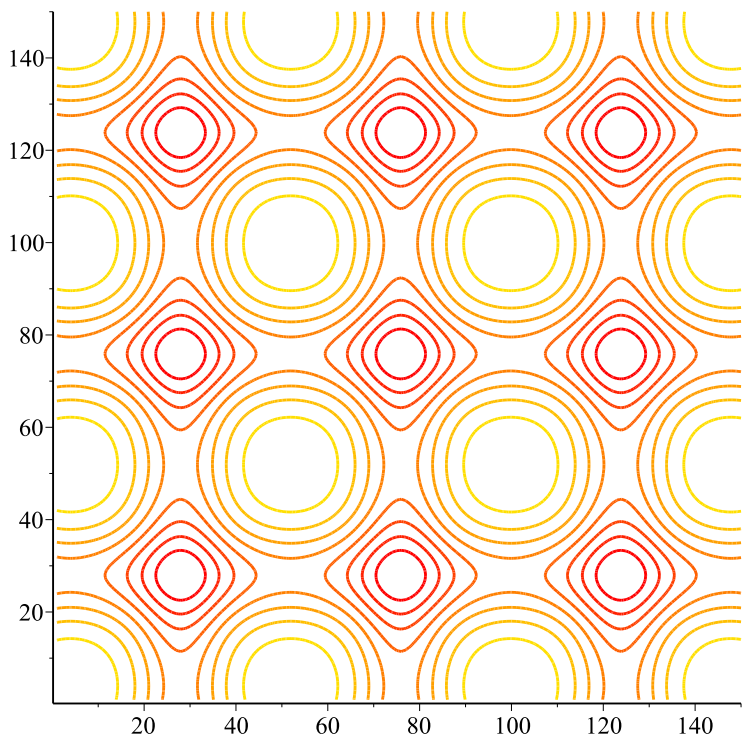}
 	\caption{ contour plot of a solution of the equation (\ref{v3}).}
 	\label{fig:6}
 \end{figure}

 Back to the Tzitz\'{e}ica equation (\ref{tsi}) and introduce a new function
$$ v = \exp(\psi) . $$
Then the equation (\ref{tsi}) is rewritten as follows
 \begin{equation} \label{v3} 
	v(v_{xx} +v_{yy}) -v^2_x-	v^2_y -v^3/2+1 = 0 .
\end{equation}
First, we look for solutions to this equation in the form
\begin{equation} \label{v4}
	v= \frac{s_0+s_1F+s_2G}{p_0+p_1F+p_2G} ,
\end{equation}
where $F$ and $G$ satisfy the equations (\ref{Fprime}), (\ref{Gprime}).
Substitute $v$ again into the right side (\ref{v3}) and repeating the reasoning above, we get the following representation
 $$ v= s_0+F+G , \qquad s_0\in\mathbb{R}.  $$
The equations for the functions $F$ and $G$ have the form
 \begin{eqnarray} 
 	(F_x^{\prime})^2= 2F^3+ \left(3s_0+\frac{a_1-b_1}{2s_0}\right)F^2+a_1F-s_0^3
 	+\frac{s_0(a_1+b_1)}{2}+1-b_0 ,\nonumber \\
 	(G_y^{\prime})^2= 2G^3 +\left(3s_0+\frac{-a_1+b_1}{2s_0}\right)G^2+b_1G +b_0 \nonumber .
 \end{eqnarray}
 with  $s_0\neq 0$. 
 The solutions of the previous two equations are expressed in terms of the Weierstrass elliptic functions.
 To get a specific solution, let's set $s_0=1, b_0=-0.1$, $a_1=b_1=-0.2$.
 As the initial data, we choose the values $F(0)=G(0)=-0.2$.
 A two-dimensional contour graph of the function $v$ is shown in Fig. \ref{fig:6}. 
 On one of the contour lines, the function $v$ is equal to zero, and therefore the stream function
  $\psi$ is not defined on it.
 
 We are now looking for a function $v$ in the form
 \begin{equation} \label{v4} 
 	v = \frac{s_0+s_1F+s_2G+s_3FG}{p_0+p_1F+p_2G+p_3FG} \ . 
 \end{equation}
By repeating the previous reasoning, we can obtain several different representations for $v$.
The simplest of them is
$$ v = \frac{p_0+ p_3FG}{s_2G} \ . $$
In this case, the functions $F$ and $G$ must satisfy the following equations
\begin{equation} \label{F3} 
	(F_x^{\prime})^2=2p_3 F^3 -b_2 F^2 +b_3p_0F/p_3 +a_0 ,
\end{equation} 
\begin{equation} \label{G3} 
	(G_y^{\prime})^2=b_4 G^4 +b_3F^3 +b_2G^2 +2p_0G/s_2 .
\end{equation}

In conclusion, we give a few more representations for the function of the function~$v$:
$$ v = \frac{p_0+p_1F+p_3FG}{s_1F} \ ,$$
$$ v =\frac{p_0+p_1F+p_0s_3G/s_1 +p_3FG}{s_1F +s_3FG}   ,\ $$
$$ v= \frac{p_0+p_1F+p_2G+p_3FG}{p_0s_3G/p_1+s_3FG}     . $$
We do not present the form of the corresponding equations (\ref{Fprime}), (\ref{Gprime}) due to their cumbersomeness.

\section{Conclusions}

In this paper, we found new classes of solutions of the two-dimensional Euler equations for an inviscid fluid.
They describe various smooth and singular vortex flows.
A new method for constructing solutions in elliptic functions is proposed.
It would be useful to try to classify all such solutions.
It is interesting to construct similar solutions in elliptic functions for other mathematical models.
The question of finding similar solutions for non-stationary Euler equations remains open.
It would be very important to prove the quantization of the flux of fluid volume across the closed curve  without the assumption of representing solutions in the form (\ref{arctan}).



\end{document}